# MoS$_2$ nanoribbons thermoelectric generators

Abbas Arab and Qiliang Li

Department of Electrical and Computer Engineering, George Mason University,

Fairfax, VA 22033

**Abstract**

In this work, we have designed and simulated new thermoelectric generator based on monolayer and few-layer MoS$_2$ nanoribbons. The proposed thermoelectric generator is composed of thermocouples made of both n-type and p-type MoS$_2$ nanoribbon legs. Density functional tight-binding non-equilibrium Green's function (DFTB-NEGF) method has been used to calculate the transmission spectrums of MoS$_2$ armchair and zigzag nanoribbons. Phonon transmission spectrums are calculated based on parameterization of Stillinger-Weber potential. Thermoelectric figure of merit, $ZT$, is calculated using these electronic and phonon transmission spectrums. Monolayer and bilayer MoS$_2$ armchair nanoribbons are found to have the highest $ZT$ value for p-type and n-type legs, respectively. Moreover, we have compared the thermoelectric current of doped monolayer MoS$_2$ armchair nanoribbons and Si thin films. Results indicate that thermoelectric current of MoS$_2$ monolayer nanoribbons is several orders of magnitude higher than that of Si thin films.

# I. Introduction

The advent of Graphene [1–3], a two-dimensional (2D) sheet of carbon atoms in honeycomb lattice, has stimulated great interest and intensive research on the properties of 2D materials. More recently, a new family of 2D semiconductor materials has been proposed, namely Transition Metal Dichalcogenides (TMDs). The presence of a bandgap in TMDs, a crucial property for microelectronics applications, has attracted much attention in comparison with the gapless Graphene. Among all TMDs, Molybdenum disulfide ($MoS_2$) is the most representative, widely interesting and intensively studied one [4–9], partially because it is relatively stable and readily available. $MoS_2$ has been used as a dry lubricant in automobile industry due to its low friction properties. Recently, it has been studied for applications in field effect transistors [10–14], photovoltaics [15] and photocatalysis [16].

In general, bulk TMDs has a layered structure. Each layer is formed by a plane of transition metal atoms sandwiched between two planes of chalcogen atoms in trigonal prismatic arrangements as illustrated in Fig. 1. Strong intralayer covalent bonding, in contrast to weak interlayer van der Waals forces [17] make it possible to fabricate high-quality monolayer $MoS_2$ by exfoliation technique [17–19]. A desirable bandgap [17–20], comparable carrier mobility with those of Si thin film and Graphene nanoribbons [10, 19, 21, 22] together with excellent thermal stability [19] and surface free from dangling bonds [23, 24], makes 2D $MoS_2$ a very attractive candidate for device applications [25–27].

Compared to the research progress in its electronic and mechanical characteristics [28–31], thermoelectric (TE) properties of $MoS_2$ have not been widely studied. Thermoelectrics provide a way of converting thermal energy into electricity [32]. Thermoelectric generator is expected to play an important role in increasing demand for clean energy in future [33]. In general, a TE generator module is made of an array of thermocouples. As illustrated in Fig. 2, each thermocouple, the basic unit of a TE generator, is made of a p-type and n-type semiconductors, named as legs, connected thermally in parallel and electrically in series. Temperature gradient across thermocouple is the driving force inducing electrical current.

The research on thermoelectric materials has been one of the major topics since 1950s when basic science of thermoelectrics was well founded [34]. *$Bi_2Te_3$* and the similar alloys have played a main role in the application of thermoelectric devices [35–37]. It is well-known that efficiency of thermoelectric conversion can be evaluated by a dimensionless figure of merit $ZT = GS^2T/(\kappa_e + \kappa_{ph})$ [34], in which $G$, $S$, $\kappa_e$, $\kappa_{ph}$ and $T$ are electrical conductance, Seebeck's coefficient, electronic contribution to thermal conductance, phonon contribution to thermal conductance and absolute temperature, respectively. In order to have a high $ZT$, it is desirable to have a high electrical conductance and large Seebeck's coefficient and low thermal conductance. These parameters mainly depend on the intrinsic properties of materials and they are generally coupled with each other. Enhancement to one of them may degrade the other and the overall effect will not change. In three decades after 1950s, only incremental progress was made due to the difficulty in fine-tuning of these parameters [33].

Recently, new wave of research on thermoelectric field has been initiated because nanoscale structures may enhance thermoelectric efficiency. It was shown that quantum confinement of charge carriers in quantum-well super-lattices [38], quantum-wires [39] as well as bulk samples containing nanostructured constituents [33] will enhance thermoelectric conversion. It is known that Density of States (DOS) of low-dimensional materials exhibits sharp changes around Fermi level [33, 38, 39]. As a result, Seebeck's coefficient, which depends on logarithmic derivative of DOS, is significantly enhanced, and hence, the ZT increases. In addition to an increase in Seebeck's coefficient, low dimensional materials are known to benefit from higher phonon scattering and consequently lower phonon thermal conductance [33]. Low phonon thermal conductivity ($\kappa_{ph}$) has been already reported for TMDs: $\kappa_{ph}$ of $MoS_2$ thin films and disordered layered $WSe_2$ are about 0.1 W/mK to 1 W/mK [40] and 0.05 W/mK [41], respectively. In addition, it has been reported that $MoS_2$ has anisotropic thermal properties [42], which provides another degree of freedom to optimize TE conversion performance.

In this work, for the first time to our knowledge, thermoelectric properties of mono-, bi-, tri- and quadlayer armchair and zigzag $MoS_2$ nanoribbons have been studied for electricity generation. It is found that as the number of layers increases from monolayer to quadlayer nanoribbons, both transmission spectrum and phonon thermal conductance increase. In addition, strong electronic and thermal transport anisotropy is found between zigzag nanoribbons (ZZNR) and armchair nanoribbons (ACNR). Transmission coefficient and phonon thermal conductance of ZZNR is higher than those of ACNR with the same number of layers. Their effect on *ZT* has been studied in this work. In addition to electronic and thermal anisotropy properties, monolayer $MoS_2$ nanoribbons show p-type behavior. The Fermi level in transmission spectrum is closer to valence band maximum. In contrast to monolayer nanoribbons, few-layer nanoribbons act more n-type as Fermi level is closer to conduction band minimum. This phenomenon is very desirable since both p-type and n-type semiconductors are required to construct TE generators. It is an advantage that the same 2D materials can be used for both p-type and n-type semiconductor legs in thermocouple. In addition, TE conversion of Si thin film TE generator with the same dimensions as $MoS_2$ nanoribbon TE generator has been studied by using Synopsys TCAD software [43]. The comparison indicates that $MoS_2$ nanoribbons exhibit much better TE conversion efficiency.

## II. Computational Method in Simulation

The computational model used in this paper is based on self-consistent density functional tight-binding non-equilibrium Green's function (DFTB-NEGF) method [44–46] implemented in QuantumWise ATK software package [47]. Prior to the calculations of carrier transport, the structure has been relaxed to optimized force and stress of 0.02 eV/Å and 0.02 eV/ Å$^3$, respectively. The relaxation calculations is implemented by using Generalized Gradient Approximation (GGA) exchange correlation with a Double Zeta Polarized (DZP) basis set and a mesh cut-off energy of 75 Ha.

Top view of nanoribbons device structures studied in this paper is illustrated in Fig. 3. Device can be divided into three regions: left, right and central. Left and right regions are the electrodes treated with semi-infinite boundary conditions. Their properties can be described by solving for the bulk material. The voltage and temperature bias are applied on electrode regions. Central region includes a repetition of each electrode region in order to screen out perturbations introduced in the scattering regions. In order to have an insight on the effect of lattice direction and thickness on the intrinsic TE properties of nanoribbons, no perturbation is introduced in the scattering region in this work.

Central region shown in Fig. 3, should be large enough to accommodate both the voltage and temperature drop within itself. Due to computational constraints, we used 149, 299, 449 and 599 atoms supercell as central region in mono-, bi-, tri- and quadlayer devices, respectively. Using infinitesimal voltage and temperature drop, i.e. working in linear regime, makes our approximation valid. In addition, a vacuum spacing of 20 Å is added to each side of the device super cell to suppress any interaction caused by periodic boundary condition at out-of-plane direction.

In order to calculate linear transport properties of the system, Landauer-Buttiker [48, 49] formula is used, in which transport coefficients are calculated from Green's function. This formulism is correct in absence of inelastic scattering and phase-changing mechanisms. DFTB-NEGF method is chosen since it is proven to be a fast and computationally efficient method for a systems containing a large number of atoms, such as nanowires and nanoribbons [50–52]. For DFTB calculations, semi-empirical Slater-Koster [53, 54] with DFTB-CP2K [55, 56] parameters available for Mo-S in Quantum Wise ATK is used. Monkhorst-Pack k-grid [57] of $1 \times 1 \times 100$ with a density mesh cut-off of 10 Ha is used for device supercell. Results of DFTB calculations using CP2K set have been shown to be in a good agreement with results of DFT calculations within Localized Density Approximation (LDA) [58].

Electrical current $I$ in the device at the linear transport regime is given by:

$$I = \frac{2q}{h} \int dE T(E)\{f_L(E, \mu_L) - f_R(E, \mu_R)\} \qquad (1)$$

where factor 2 counts for spin degeneracy, $q$ is electrical charge of carrier, $h$ is Planck's constant, $T(E)$ is transmission spectrum of device, $\mu_{L(R)}$ is chemical potential of left (right) electrodes and $f_{L(R)}(E, \mu_{L(R)})$ is the Fermi distribution of left (right) electrode. In linear response regime, it is assumed that $\Delta\mu = \mu_L - \mu_R$ and $\Delta T = T_L - T_R$ are infinitesimally small. As a result, Equation (1) will be reduced to:

$$I = \frac{2q}{h} \int dE \cdot T(E) \cdot \left\{-\frac{\partial f(E,\mu)}{\partial E}\right\} \tag{2}$$

Electronic contribution to TE properties, which is including electrical conductance ($G$), Seebeck's coefficient ($S$) and electronic thermal conductance ($\kappa_e$), can be calculated by using the followings:

$$G = qL_0 \tag{3}$$

$$S = \frac{L_1}{qTL_0} \tag{4}$$

$$\kappa_e = \frac{1}{T}\left(L_2 - \frac{L_1^2}{L_0}\right) \tag{5}$$

where $L_n$ is expressed as:

$$L_n = \frac{2}{h} \int_{-\infty}^{\infty} dE \cdot T(E) \cdot \left(-\frac{\partial f(E,\mu)}{\partial E}\right) \cdot (E - \mu)^n. \tag{6}$$

Phonon transmission spectrum is calculated based on parameterization of Stillinger-Weber [59] potential for MoS$_2$ [60] as implemented in Quantum Wise ATK package. Phonon thermal conductance ($\kappa_{ph}$) can be calculated as:

$$\kappa_{ph} = \lim_{\Delta T \to 0} \frac{\frac{1}{h}\int_0^\infty dE \cdot T_{ph}(E) \cdot E \cdot (B(E,T_L) - B(E,T_R))}{\Delta T} \tag{7}$$

where $T_{ph}(E)$ is phonon transmission spectrum; $B(E, T_L)$ and $B(E, T_R)$ are Bose-Einstein distribution of the left and right electrodes, respectively; $T_{L(R)}$ is temperature of left(right) electrode and $E$ is energy of phonon. In linear response regime, $\Delta T = T_L - T_R \approx 0$ and equation (7) becomes:

$$\kappa_{ph} = \frac{1}{h} \int_0^\infty dE \cdot T_{ph}(E) \cdot E \cdot \left(-\frac{\partial B(E,T)}{\partial T}\right). \tag{8}$$

It is worth mentioning that the phonon thermal conductance calculations in this paper are performed in the absence of any phonon decaying mechanisms. Hence, the calculations set the upper limit for phonon thermal conductance of pure MoS2. In reality however, there would be a few mechanisms which tend to suppress phonon conduction such as rough surface, edge imperfectness of ribbons, scattering centers, etc. ZT figure of merit calculated in this paper, therefore is the minimum of what actually can be achieved by these materials. TE figure of merit is calculated by using the above information:

$$ZT = \frac{GS^2T}{\kappa_e + \kappa_{ph}}. \tag{9}$$

## III. Results and Discussion

Transmission spectrum characterizes the electrical behavior of the proposed devices. Electrical factors that affect TE figure of merit include electrical conductance ($G$), electronic thermal conductance ($\kappa_e$) and Seebeck's coefficient ($S$). These factors can be derived from transmission spectrum as described in the previous section. Transmission spectrums for monolayer and few-layer MoS$_2$ ACNR and ZZNR are illustrated in Fig. 4. Fermi level for pristine nanoribbons is located at $E - E_f = 0\ eV$ and is closer to Valence Band Maximum (VBM) in monolayer nanoribbons while for few-layer nanoribbons is closer to Conduction Band Minimum (CBM). In other words, monolayer nanoribbons act more as p-type while few-layer nanoribbons act more as n-type. Few-layer nanoribbons have similar transmission spectrum profile which is totally different than those of monolayer nanoribbons. This inhomogeneity of transmission spectrums is due to absence of interlayer van der Waals forces in monolayer nanoribbons. Further study of Fig. 4 indicates that as the number of layers increases from two to four layers, the profile of transmission spectrum does not change. However, the amplitude of transmission spectrum increases, indicating that each layer provides an independent channel to conduct carriers [61]. Furthermore, ZZNR is found to have higher transmission amplitude in comparison with ACNR. It is expected to be more conductive than ACNR consequently.

In semiconducting materials, phonon thermal conductance ($\kappa_{ph}$) is several times larger than $\kappa_e$ and outplays the impact of $\kappa_e$ on TE figure of merit. $\kappa_{ph}$ of monolayer and few-layer ACNR and ZZNR vs. temperature are illustrated in Fig. 5. As shown in Fig. 5, $\kappa_{ph}$ is almost independent of temperature. It is closely a constant in a wide range of temperatures (from 200K to 500K). In addition, ZZNR shows larger $\kappa_{ph}$ than ACNR as was pointed out by Jiang [42] due to the alignment of one vibrational mode in transport direction along zigzag orientation. These results also suggest that $\kappa_{ph}$ of both ZZNR and ACNR increases as the number of layers increases. The rate of increase in $\kappa_{ph}$ is more in ZZNR than in ACNR. Our results of $\kappa_{ph}$ for monolayer MoS$_2$ is in a good agreement with findings by Huang [62].

From factors playing role in TE figure of merit, $G$ and $\kappa_e$ follow the profile of transmission spectrum, i.e. as the Fermi level moves into valence or conduction bands, transmission increases, and hence, there are more carriers to be conducted both thermally and electrically. In contrast to $G$ and $\kappa_e$, it is typical for semiconductor materials that Seebeck's coefficient ($S$) decreases as Fermi level moves into valence and conduction bands. Therefore $G$ and $S$ are competing with each other and their product in the form of $S^2G$, known as power factor, reaches its maximum at an optimum position of Fermi energy [33, 36, 62].

$ZT$ values of monolayer and few-layer MoS$_2$ ACNR and ZZNR vs. Fermi level position at four temperatures are illustrated in Fig. 6. There are two main peaks in $ZT$, separated by a bandgap, corresponding to valence band and conduction band. VBM and CBM are specified in each plot by vertical dashed lines. Fermi level position for pristine nanoribbons is at $E - E_f = 0$. In this study, TE figure of merit is referred to as ZT of n-doped or ZT of p-doped as Fermi level is approaching conduction band or valence band, respectively. It is shown in Fig. 6 that for all monolayer and few-layer nanoribbons, $ZT$ values of n-doped nanoribbons are higher than those of p-doped.

As temperature increases, amplitude of $ZT$ also increases since $ZT$ is proportional to the temperature. In addition, rising temperature broadens Fermi distribution. This broadening will populate states in energies higher than Fermi level, which were unpopulated in lower temperatures. These newly occupied states contribute to both electrical and thermal conduction. It means that electrical conductance increases in energies which has insignificant contribution to conduction in lower temperatures, resulting in broadening of $ZT$ peaks vs. energy. Further study of Fig. 6 shows that in $ZT$ of n-doped few-layer ZZNR, there are two peaks and as temperature increases the second peak merges into the first peak. This pattern is not seen in $ZT$ of n-doped few-layer ACNR. This behavior can be attributed to the narrower peak in few-layer ZZNR transmission spectrum in conduction band, compared to that of ACNR seen in Fig. 4. Transmission spectrums for both few-layer ACNR and ZZNR in conduction band are composed of two peaks. Fall of transmission spectrum in conduction band, between these two peaks, is sharper for ZZNR compared to that of ACNR. As Fermi level approaches the point at which transmission spectrum falls, $ZT$ also decreases due to decrease in transmission modes. This phenomenon is more dramatic in lower temperatures since in higher temperatures, as described above, broadening of Fermi distribution compensates lack of transmission modes by populating states in higher energies. As a result, by increasing temperature, two $ZT$ peaks merge together. Broader transmission spectrum of few-layer ACNR in conduction band compared to that of ZZNR, makes this double-peak characteristic in $ZT$ profile far less pronounced, even in low temperatures.

Peak values of $ZT$ for p-doped and n-doped monolayer and few-layer ACNR and ZZNR vs. temperature are shown in Fig. 7. As it was expected from Eq. (9), $ZT$ is quite linear with temperature. Monolayer nanoribbons have higher $ZT$ in both n-doped and p-doped plots compared to few-layer nanoribbons. $ZT$ values are larger than unity in both n-doped monolayer ACNR and ZZNR at $T = 500K$. However, it was shown by transmission spectrum that monolayer nanoribbons are more p-type than n-type. More doping is needed to shift the Fermi level close to conduction band rather than valence band and hence take advantage of high $ZT$ values of n-doped monolayer nanoribbons. Therefore, for n-type leg in thermocouple, few-layer nanoribbons are more preferred. Results in Fig. 7 show that bilayer ACNR exhibits the highest n-doped $ZT$ compared to the other few-layer structures and is the best material to be implemented as n-type leg in thermocouples.

As discussed previously, in order to take advantage of the highest $ZT$ value possible, $MoS_2$ nanoribbons should be doped in order to shift Fermi level to energy of peak values of $ZT$ profile. Substitutional doping of TMD samples has been observed experimentally under exposure to 80keV electron beam irradiation [63]. Also, a first principal study of effect of this doping approach for transition metal dopants as well as non-metal dopants is reported in [64, 65]. In order to examine the TE current of MoS$_2$ nanoribbons, we have simulated a monolayer ACNR doped with various dopant species. We followed the same substitutional approach for doping our ACNR. Transition metal atoms (Re, Ru and Ta) are used as the replacing dopants for Molybdenum, and non-transition metal atoms (As, Br, Cl and P) are used for Sulphur [66]. In order to screen out the perturbation caused by doping properly, only one dopant atom was inserted in central region of device. A temperature gradient has been set across the nanoribbon by fixing the temperature of right electrode to $T = 300\ K$ and changing temperature of left electrode from $T = 250K$ to $T = 350K$ (for device configuration, see Fig. 3). Results are shown in Fig. 8. TE current of ACNR shows strong dependence on the type of dopant atom. While Arsenic does not show any effect on thermoelectric current, $P$ and $Ta$ showed a similar boost to current. For n-type dopant, $Ru$ exhibits the best current boost in comparison with other dopants. It should be noted that doping in MoS$_2$ monolayer at nanoscale will induce device to device performance variation [67].

These results are compared with TE current of Si thin film doped with acceptor (B) concentration of $N_A = 1 \times 10^{16}\ cm^{-3}$ with various film thicknesses (also shown in Fig. 8). For Si thin film with film thickness of $t = 5nm$, TE current density reaches $J \approx 0.001\ A/cm$ at $\Delta T = 50K$. In comparison, monolayer $Ru$-doped MoS$_2$ ACNR has TE current density $\approx 0.2\ A/cm$ at $\Delta T = 50K$, more than two orders of magnitude larger. Decreasing thickness of Si film makes them more resistive and TE current decreases consequently, as shown in Fig.8. Superiority of MoS$_2$-based thermocouples will be more dramatic if we compare its TE performance with those of thinner Si films, especially $1nm$-thick Si films which is almost the same thickness of monolayer MoS$_2$.

Thermocouples, as was mentioned in previous section, are made of both p-type and n-type semiconductors. In order to compare the performance of monolayer MoS$_2$-based and Si-based thermocouples, TE current of both of these materials is illustrated in Fig. 9. For Si-based thermocouples, p-doped (B) and n-doped (As) films with thickness of $t = 5nm$ and with doping concentration of $N_{A,D} = 1 \times 10^{16}\ cm^{-3}$ is used. For monolayer MoS$_2$ TE conversion, Ru-doped and P-doped nanoribbons are the best n-type and p-type nanoribbons, respectively. These two nanoribbons are chosen to construct the thermocouple based on monolayer MoS$_2$. Fig. 9 shows that thermocouples based on monolayer MoS$_2$ are far more superior than thermocouples based on Si thin films.

## IV. Conclusion

In summary, we proposed a TE generator based on monolayer and few-layer MoS$_2$ nanoribbons. In order to find the optimum structure for the proposed thermocouple, first-principle simulation has been performed to calculate TE figure of merit $ZT$ for monolayer and few-layer MoS$_2$ ACNR and ZZNR. It was shown that in monolayer MoS$_2$ nanoribbons, Fermi level is closer to valence band in contrast to few-layer nanoribbons where it is closer to conduction band. This behavior is desirable since the same 2D material can be used as p-type and n-type semiconducting leg in thermocouple, hence simplifying fabrication process. Monolayer MoS$_2$ ACNR is shown to have the highest *ZT* value as p-type semiconducting leg, while among few-layer nanoribbons, bilayer MoS$_2$ ACNR shows the highest *ZT* value as n-type semiconductor leg. Moreover, compared to Si films, MoS$_2$ monolayer nanoribbons are two orders better in achieving higher TE current.

References:

[1] K. S. Novoselov, D. Jiang, F. Schedin, T. J. Booth, V. V. Khotkevich, S. V. Morozov, and A. K. Geim, "Two-dimensional atomic crystals," *Proc. Natl. Acad. Sci. U. S. A.*, vol. 102, no. 30, pp. 10451–10453, 2005.

[2] K. S. A. Novoselov, A. K. Geim, Sv. Morozov, D. Jiang, M. K. I. Grigorieva, S. V. Dubonos, and A. A. Firsov, "Two-dimensional gas of massless Dirac fermions in graphene," *nature*, vol. 438, no. 7065, pp. 197–200, 2005.

[3] A. K. Geim and K. S. Novoselov, "The rise of graphene," *Nat. Mater.*, vol. 6, no. 3, pp. 183–191, 2007.

[4] Y. Li, Z. Zhou, S. Zhang, and Z. Chen, "MoS2 nanoribbons: High stability and unusual electronic and magnetic properties," *J. Am. Chem. Soc.*, vol. 130, no. 49, pp. 16739–16744, 2008.

[5] C. Ataca, H. Sahin, E. Akturk, and S. Ciraci, "Mechanical and electronic properties of MoS2 nanoribbons and their defects," *J. Phys. Chem. C*, vol. 115, no. 10, pp. 3934–3941, 2011.

[6] H. Pan and Y.-W. Zhang, "Tuning the electronic and magnetic properties of MoS2 nanoribbons by strain engineering," *J. Phys. Chem. C*, vol. 116, no. 21, pp. 11752–11757, 2012.

[7] L. Kou, C. Tang, Y. Zhang, T. Heine, C. Chen, and T. Frauenheim, "Tuning Magnetism and Electronic Phase Transitions by Strain and Electric Field in Zigzag MoS2 Nanoribbons," *J. Phys. Chem. Lett.*, vol. 3, no. 20, pp. 2934–2941, 2012.

[8] A. R. Botello-Mendez, F. Lopez-Urias, M. Terrones, and H. Terrones, "Metallic and ferromagnetic edges in molybdenum disulfide nanoribbons," *Nanotechnology*, vol. 20, no. 32, p. 325703, 2009.

[9] K. Dolui, C. D. Pemmaraju, and S. Sanvito, "Electric field effects on armchair MoS2 nanoribbons," *Acs Nano*, vol. 6, no. 6, pp. 4823–4834, 2012.

[10] S. Ghatak, A. N. Pal, and A. Ghosh, "Nature of electronic states in atomically thin MoS2 field-effect transistors," *Acs Nano*, vol. 5, no. 10, pp. 7707–7712, 2011.

[11] D. J. Late, B. Liu, H. R. Matte, V. P. Dravid, and C. N. R. Rao, "Hysteresis in single-layer MoS2 field effect transistors," *Acs Nano*, vol. 6, no. 6, pp. 5635–5641, 2012.

[12] H. Qiu, L. Pan, Z. Yao, J. Li, Y. Shi, and X. Wang, "Electrical characterization of back-gated bi-layer MoS2 field-effect transistors and the effect of ambient on their performances," *Appl. Phys. Lett.*, vol. 100, no. 12, p. 123104, 2012.

[13] W. Bao, X. Cai, D. Kim, K. Sridhara, and M. S. Fuhrer, "High mobility ambipolar MoS2 field-effect transistors: Substrate and dielectric effects," *Appl. Phys. Lett.*, vol. 102, no. 4, p. 042104, 2013.

[14] H. Li, Z. Yin, Q. He, H. Li, X. Huang, G. Lu, D. W. H. Fam, A. I. Y. Tok, Q. Zhang, and H. Zhang, "Fabrication of Single-and Multilayer MoS2 Film-Based Field-Effect Transistors for Sensing NO at Room Temperature," *Small*, vol. 8, no. 1, pp. 63–67, 2012.

[15] E. Gourmelon, O. Lignier, H. Hadouda, G. Couturier, J. C. Bernede, J. Tedd, J. Pouzet, and J. Salardenne, "MS$_2$(M= W, Mo) photosensitive thin films for solar cells," *Sol. Energy Mater. Sol. Cells*, vol. 46, no. 2, pp. 115–121, 1997.

[16] X. Zong, H. Yan, G. Wu, G. Ma, F. Wen, L. Wang, and C. Li, "Enhancement of photocatalytic H2 evolution on CdS by loading MoS2 as cocatalyst under visible light irradiation," *J. Am. Chem. Soc.*, vol. 130, no. 23, pp. 7176–7177, 2008.

[17] K. F. Mak, C. Lee, J. Hone, J. Shan, and T. F. Heinz, "Atomically thin MoS 2: a new direct-gap semiconductor," *Phys. Rev. Lett.*, vol. 105, no. 13, p. 136805, 2010.

[18] A. Splendiani, L. Sun, Y. Zhang, T. Li, J. Kim, C.-Y. Chim, G. Galli, and F. Wang, "Emerging photoluminescence in monolayer MoS2," *Nano Lett.*, vol. 10, no. 4, pp. 1271–1275, 2010.

[19] B. Radisavljevic, A. Radenovic, J. Brivio, V. Giacometti, and A. Kis, "Single-layer MoS2 transistors," *Nat. Nanotechnol.*, vol. 6, no. 3, pp. 147–150, 2011.
[20] K. K. Kam and B. A. Parkinson, "Detailed photocurrent spectroscopy of the semiconducting group VIB transition metal dichalcogenides," *J. Phys. Chem.*, vol. 86, no. 4, pp. 463–467, 1982.
[21] L. Liu, S. Bala Kumar, Y. Ouyang, and J. Guo, "Performance limits of monolayer transition metal dichalcogenide transistors," *Electron Devices IEEE Trans. On*, vol. 58, no. 9, pp. 3042–3047, 2011.
[22] Y. Yoon, K. Ganapathi, and S. Salahuddin, "How good can monolayer MoS2 transistors be?," *Nano Lett.*, vol. 11, no. 9, pp. 3768–3773, 2011.
[23] E. Gourmelon, J. C. Bernede, J. Pouzet, and S. Marsillac, "Textured MoS2 thin films obtained on tungsten: Electrical properties of the W/MoS2 contact," *J. Appl. Phys.*, vol. 87, no. 3, pp. 1182–1186, 2000.
[24] A. Rothschild, S. R. Cohen, and R. Tenne, "WS2 nanotubes as tips in scanning probe microscopy," *Appl. Phys. Lett.*, vol. 75, no. 25, pp. 4025–4027, 1999.
[25] B. Radisavljevic, M. B. Whitwick, and A. Kis, "Integrated circuits and logic operations based on single-layer MoS2," *Acs Nano*, vol. 5, no. 12, pp. 9934–9938, 2011.
[26] H. Wang, L. Yu, Y.-H. Lee, Y. Shi, A. Hsu, M. L. Chin, L.-J. Li, M. Dubey, J. Kong, and T. Palacios, "Integrated circuits based on bilayer MoS2 transistors," *Nano Lett.*, vol. 12, no. 9, pp. 4674–4680, 2012.
[27] Z. Yin, H. Li, H. Li, L. Jiang, Y. Shi, Y. Sun, G. Lu, Q. Zhang, X. Chen, and H. Zhang, "Single-layer MoS2 phototransistors," *ACS Nano*, vol. 6, no. 1, pp. 74–80, 2011.
[28] S. Yu, H. D. Xiong, K. Eshun, H. Yuan, and Q. Li, "Phase transition, effective mass and carrier mobility of MoS 2 monolayer under tensile strain," *Appl. Surf. Sci.*, vol. 325, pp. 27–32, 2015.
[29] A. Kuc, N. Zibouche, and T. Heine, "Influence of quantum confinement on the electronic structure of the transition metal sulfide T S 2," *Phys. Rev. B*, vol. 83, no. 24, p. 245213, 2011.
[30] R. Coehoorn, C. Haas, J. Dijkstra, C. J. F. Flipse, R. A. De Groot, and A. Wold, "Electronic structure of MoSe 2, MoS 2, and WSe 2. I. Band-structure calculations and photoelectron spectroscopy," *Phys. Rev. B*, vol. 35, no. 12, p. 6195, 1987.
[31] S. Lebegue and O. Eriksson, "Electronic structure of two-dimensional crystals from ab initio theory," *Phys. Rev. B*, vol. 79, no. 11, p. 115409, 2009.
[32] F. J. DiSalvo, "Thermoelectric cooling and power generation," *Science*, vol. 285, no. 5428, pp. 703–706, 1999.
[33] M. S. Dresselhaus, G. Chen, M. Y. Tang, R. G. Yang, H. Lee, D. Z. Wang, Z. F. Ren, J.-P. Fleurial, and P. Gogna, "New Directions for Low-Dimensional Thermoelectric Materials," *Adv. Mater.*, vol. 19, no. 8, pp. 1043–1053, 2007.
[34] H. J. Goldsmid, "Basic Principles," in *Thermoelectric Refrigeration*, Springer, 1964, pp. 1–11.
[35] F. D. Rosi, "Thermoelectricity and thermoelectric power generation," *Solid-State Electron.*, vol. 11, no. 9, pp. 833–868, 1968.
[36] D. M. Rowe, *Thermoelectrics handbook: macro to nano*. CRC press, 2005.
[37] H. J. Goldsmid and R. W. Douglas, "The use of semiconductors in thermoelectric refrigeration," *Br. J. Appl. Phys.*, vol. 5, no. 11, p. 386, 1954.
[38] L. D. Hicks and M. S. Dresselhaus, "Effect of quantum-well structures on the thermoelectric figure of merit," *Phys. Rev. B*, vol. 47, no. 19, p. 12727, 1993.
[39] L. D. Hicks and M. S. Dresselhaus, "Thermoelectric figure of merit of a one-dimensional conductor," *Phys. Rev. B*, vol. 47, no. 24, p. 16631, 1993.
[40] C. Chiritescu, D. G. Cahill, N. Nguyen, D. Johnson, A. Bodapati, P. Keblinski, and P. Zschack, "Ultralow thermal conductivity in disordered, layered WSe2 crystals," *Science*, vol. 315, no. 5810, pp. 351–353, 2007.

[41] V. Varshney, S. S. Patnaik, C. Muratore, A. K. Roy, A. A. Voevodin, and B. L. Farmer, "MD simulations of molybdenum disulphide (MoS$_2$): Force-field parameterization and thermal transport behavior," *Comput. Mater. Sci.*, vol. 48, no. 1, pp. 101–108, 2010.

[42] J.-W. Jiang, X. Zhuang, and T. Rabczuk, "Orientation Dependent Thermal Conductance in Single-Layer MoS2," *Sci. Rep.*, vol. 3, 2013.

[43] "Synopsys TCAD," *Synopsys*. [Online]. Available: http://www.synopsys.com/Tools/TCAD/Pages/default.aspx. [Accessed: 07-Dec-2014].

[44] K. Stokbro, J. Taylor, M. Brandbyge, and H. Guo, "Ab-initio non-equilibrium Green's function formalism for calculating electron transport in molecular devices," in *Introducing Molecular Electronics*, Springer, 2005, pp. 117–151.

[45] M. Brandbyge, J.-L. Mozos, P. Ordejón, J. Taylor, and K. Stokbro, "Density-functional method for nonequilibrium electron transport," *Phys. Rev. B*, vol. 65, no. 16, p. 165401, 2002.

[46] H. Haug and A.-P. Jauho, *Quantum kinetics in transport and optics of semiconductors*, vol. 123. Springer, 2007.

[47] "Quantum Wise ATK Software Package." [Online]. Available: http://www.quantumwise.com/. [Accessed: 07-Nov-2014].

[48] M. Büttiker, Y. Imry, R. Landauer, and S. Pinhas, "Generalized many-channel conductance formula with application to small rings," *Phys. Rev. B*, vol. 31, no. 10, p. 6207, 1985.

[49] M. Büttiker, "Four-terminal phase-coherent conductance," *Phys. Rev. Lett.*, vol. 57, no. 14, p. 1761, 1986.

[50] I. Popov, G. Seifert, and D. Tománek, "Designing electrical contacts to MoS 2 monolayers: a computational study," *Phys. Rev. Lett.*, vol. 108, no. 15, p. 156802, 2012.

[51] G. Seifert, H. Terrones, M. Terrones, G. Jungnickel, and T. Frauenheim, "Structure and electronic properties of MoS 2 nanotubes," *Phys. Rev. Lett.*, vol. 85, no. 1, p. 146, 2000.

[52] E. Erdogan, I. H. Popov, A. N. Enyashin, and G. Seifert, "Transport properties of MoS 2 nanoribbons: edge priority," *Eur. Phys. J. B-Condens. Matter Complex Syst.*, vol. 85, no. 1, pp. 1–4, 2012.

[53] G. F. Koster and J. C. Slater, "Wave functions for impurity levels," *Phys. Rev.*, vol. 95, no. 5, p. 1167, 1954.

[54] K. Stokbro, D. E. Petersen, S. Smidstrup, A. Blom, M. Ipsen, and K. Kaasbjerg, "Semiempirical model for nanoscale device simulations," *Phys. Rev. B*, vol. 82, no. 7, p. 075420, 2010.

[55] "CP2K Consortium." [Online]. Available: http://www.cp2k.org/.

[56] "hotbit Consortium." [Online]. Available: https://trac.cc.jyu.fi/projects/hotbit.

[57] H. J. Monkhorst and J. D. Pack, "Special points for Brillouin-zone integrations," *Phys. Rev. B*, vol. 13, no. 12, p. 5188, 1976.

[58] A. Sengupta and S. Mahapatra, "Negative differential resistance and effect of defects and deformations in MoS2 armchair nanoribbon metal-oxide-semiconductor field effect transistor," *J. Appl. Phys.*, vol. 114, no. 19, p. 194513, 2013.

[59] F. H. Stillinger and T. A. Weber, "Computer simulation of local order in condensed phases of silicon," *Phys. Rev. B*, vol. 31, no. 8, p. 5262, 1985.

[60] J.-W. Jiang, H. S. Park, and T. Rabczuk, "Molecular dynamics simulations of single-layer molybdenum disulphide (MoS2): Stillinger-Weber parametrization, mechanical properties, and thermal conductivity," *J. Appl. Phys.*, vol. 114, no. 6, p. 064307, 2013.

[61] S. Das and J. Appenzeller, "Screening and interlayer coupling in multilayer MoS2," *Phys. Status Solidi RRL-Rapid Res. Lett.*, vol. 7, no. 4, pp. 268–273, 2013.

[62] W. Huang, H. Da, and G. Liang, "Thermoelectric performance of MX2 (M= Mo, W; X= S, Se) monolayers," *J. Appl. Phys.*, vol. 113, no. 10, p. 104304, 2013.

[63] H.-P. Komsa, J. Kotakoski, S. Kurasch, O. Lehtinen, U. Kaiser, and A. V. Krasheninnikov, "Two-dimensional transition metal dichalcogenides under electron irradiation: defect production and doping," *Phys. Rev. Lett.*, vol. 109, no. 3, p. 035503, 2012.

[64] C. Ataca and S. Ciraci, "Functionalization of single-layer MoS2 honeycomb structures," *J. Phys. Chem. C*, vol. 115, no. 27, pp. 13303–13311, 2011.

[65] K. Dolui, I. Rungger, C. D. Pemmaraju, and S. Sanvito, "Possible doping strategies for MoS 2 monolayers: An ab initio study," *Phys. Rev. B*, vol. 88, no. 7, p. 075420, 2013.

[66] Q. Yue, S. Chang, S. Qin, and J. Li, "Functionalization of monolayer MoS 2 by substitutional doping: a first-principles study," *Phys. Lett. A*, vol. 377, no. 19, pp. 1362–1367, 2013.

[67] K. Eshun, H. D. Xiong, S. Yu, and Q. Li, "Doping induces large variation in the electrical properties of MoS 2 monolayers," *Solid-State Electron.*, vol. 106, pp. 44–49, 2015.

Figures and Captions:

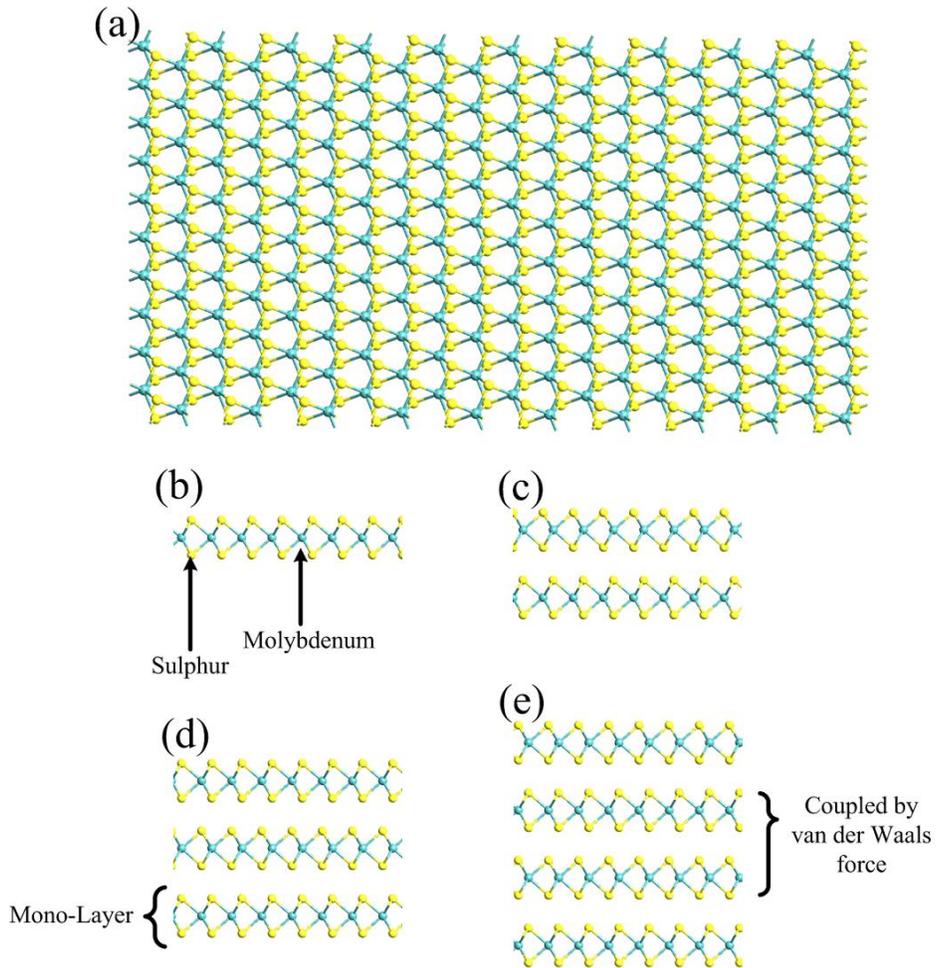

Fig. 1 Atomic structure of $MoS_2$; (a) monolayer of $MoS_2$ is made of a honeycomb sheet of Molybdenum atoms covalently sandwiched between two honeycomb sheets of Sulphur atoms. Bulk of $MoS_2$ is formed by monolayers stacked and held on top of each other by van der Waals forces. Side view of mono-, bi-, tri- and quadlayer is illustrated in parts (b), (c), (d) and (e), respectively.

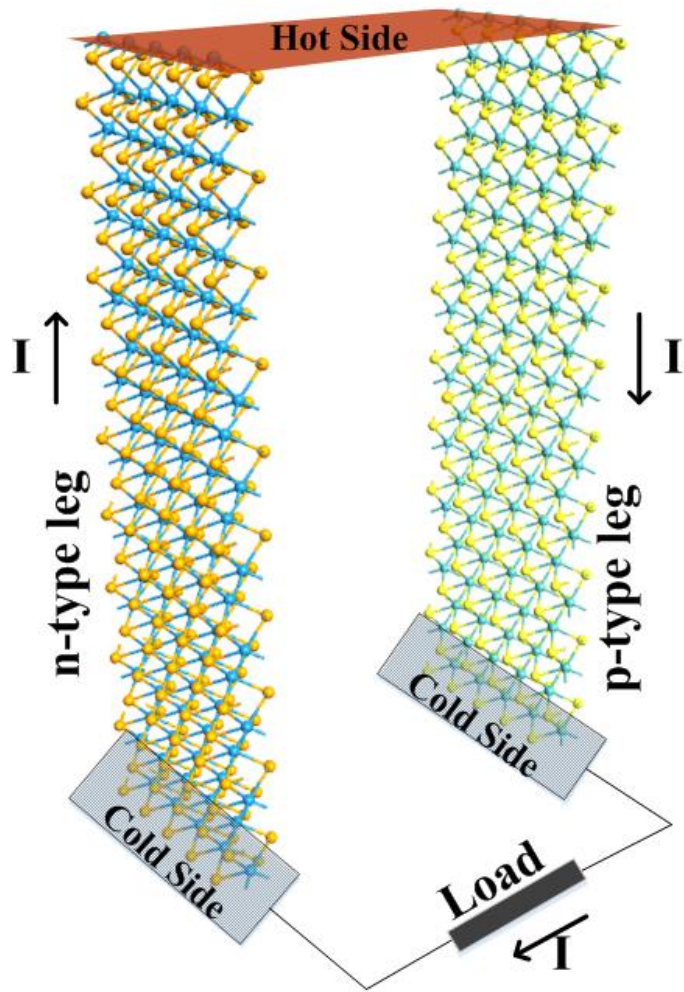

Fig. 2 Structure of proposed thermoelectric generator based on monolayer MoS$_2$. It is composed of a p-type and an n-type semiconductor, known as legs. Temperature gradient across thermocouples will induce an electrical current through thermocouple based on thermoelectric phenomena.

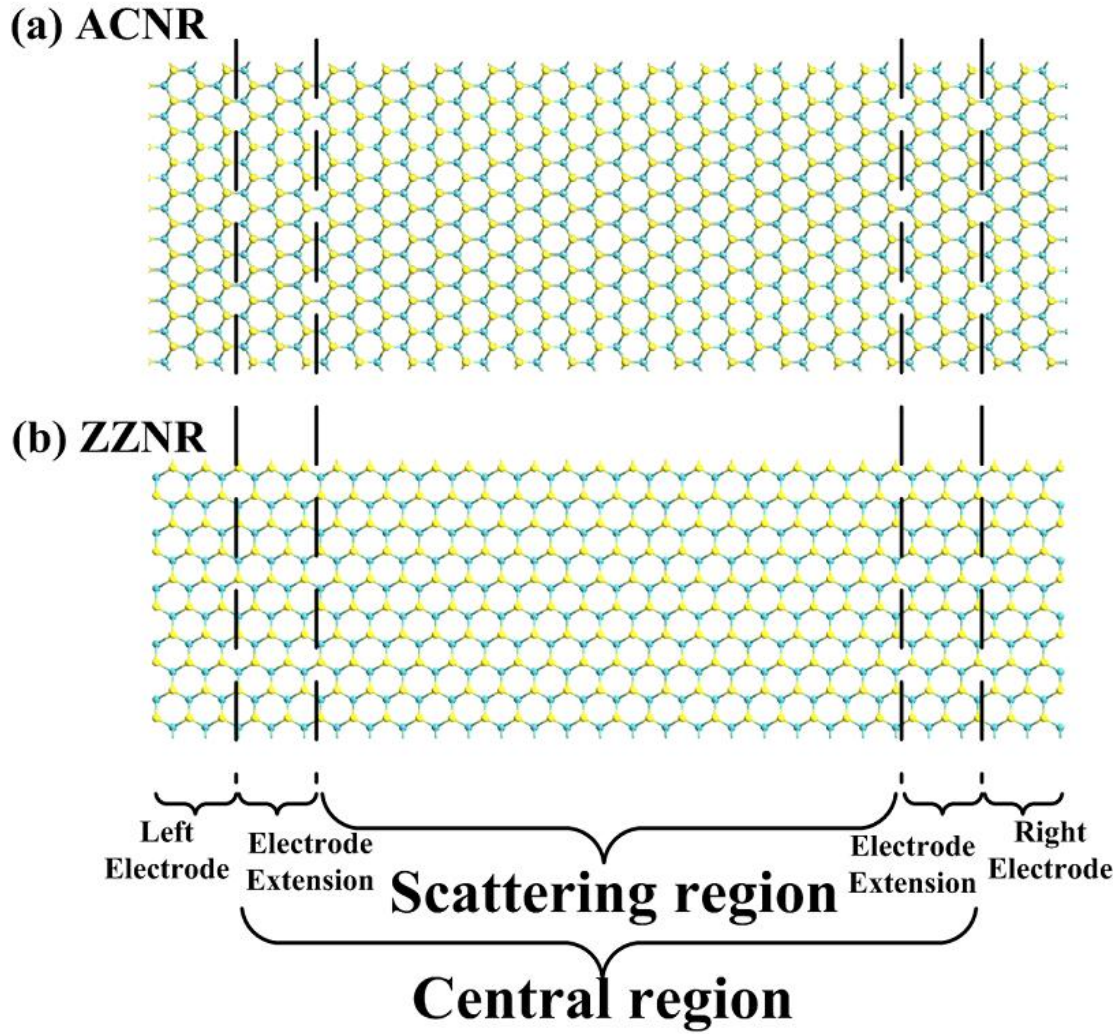

Fig. 3. Structure of a) armchair and b) zigzag nanoribbons devices studied in this paper. Each device is comprised of three regions: left electrode, central region and right electrode. Central region, itself, contains an extension of electrode regions on both sides and scattering region in the middle. Electrode regions are treated semi-infinitely. Their properties are computed by solving for bulk material. Temperature gradient is biased on electrode regions. Extension of electrode regions in central region, are used to screen out any perturbation introduced in scattering region.

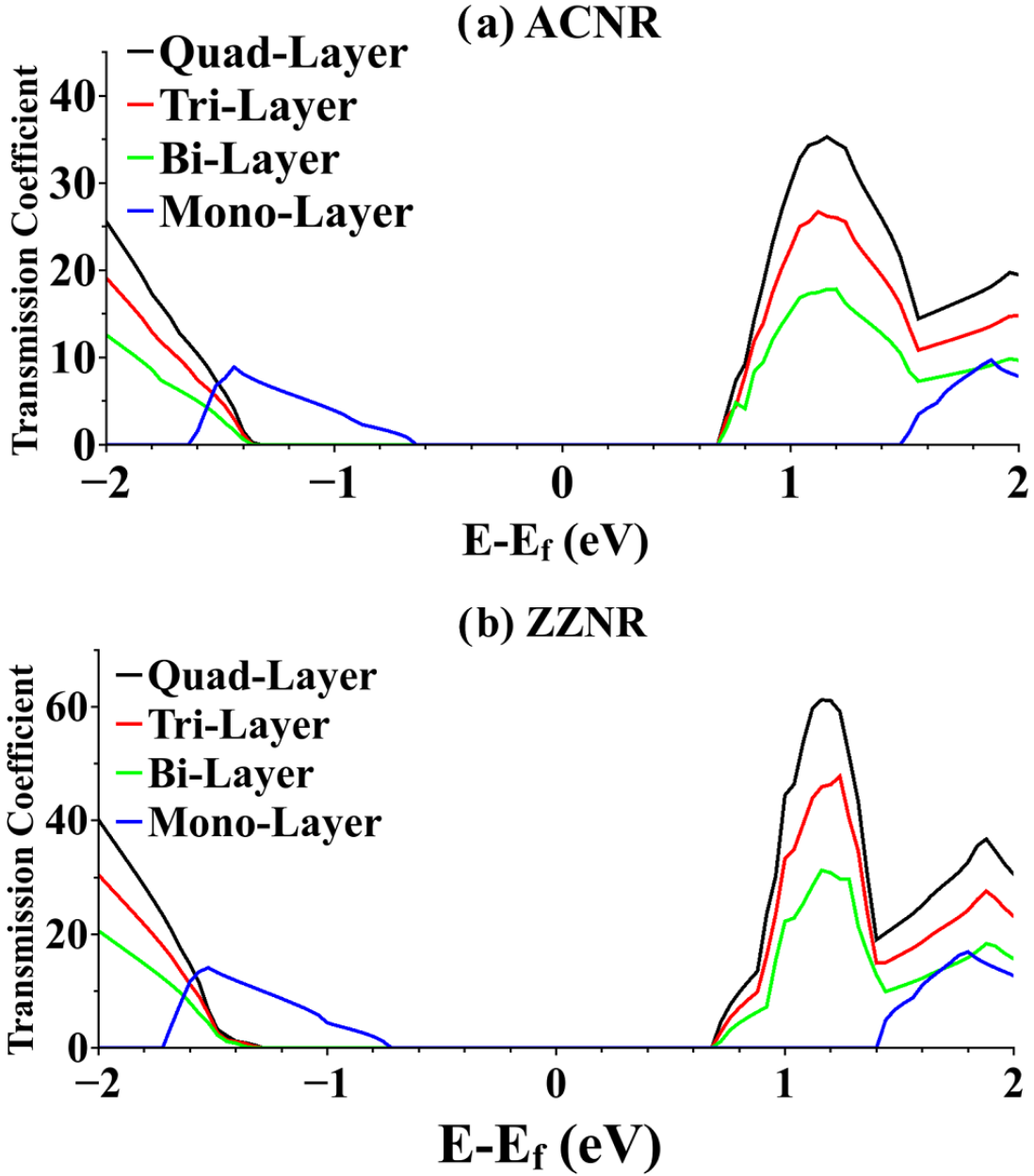

Fig. 4. Transmission spectrum for (a) ACNR and (b) ZZNR for mono-, bi-, tri- and quadlayer $MoS_2$ calculated based on DFTB-NEGF method. Fermi level is closer to valence band maximum in monolayer nanoribbons and is closer to conduction band minimum in few-layer nanoribbons. Transmission profile of monolayer and multilayer $MoS_2$ nanoribbons are different from each other due to absence of interlayer van-der-Waals forces in monolayer nanoribbons.

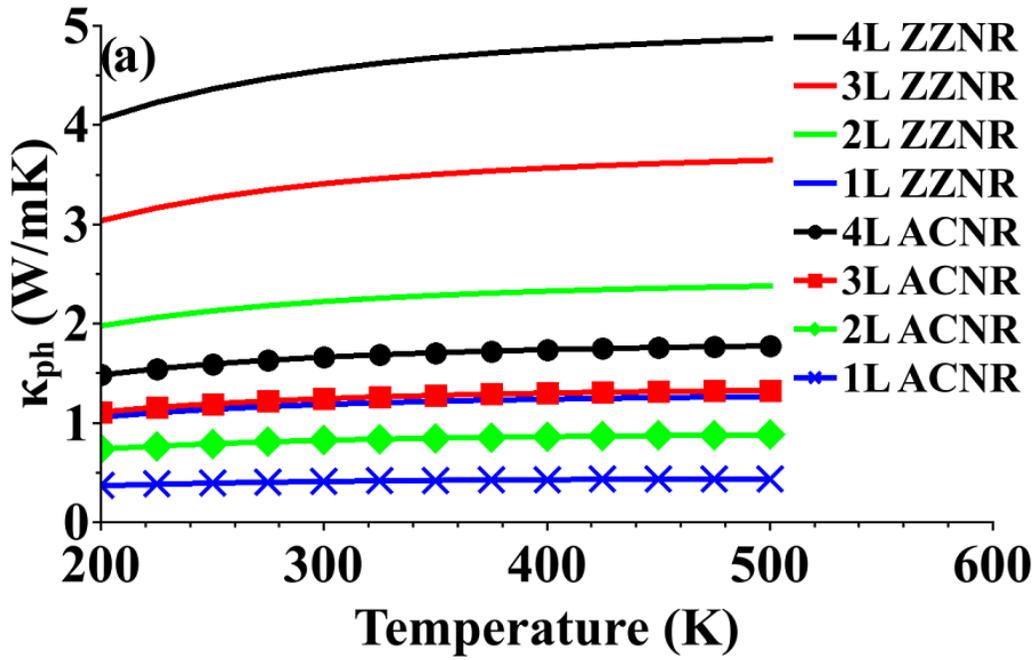

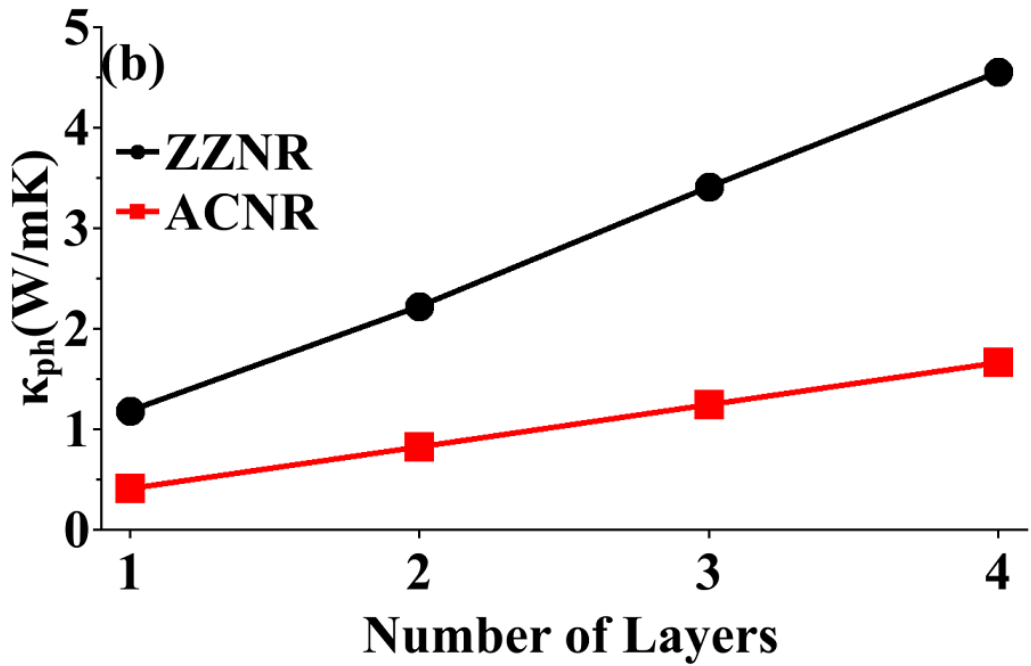

Fig. 5. (a) $\kappa_{ph}$ vs. temperature for monolayer and few-layer MoS$_2$ ACNR and ZZNR. (b) $\kappa_{ph}$ vs. number of layers for ACNR and ZZNR at $T = 300K$. $\kappa_{ph}$ for ZZNR shows higher values and greater rate of increase as number of layers increases from monolayer to quadlayer than those for ACNR.

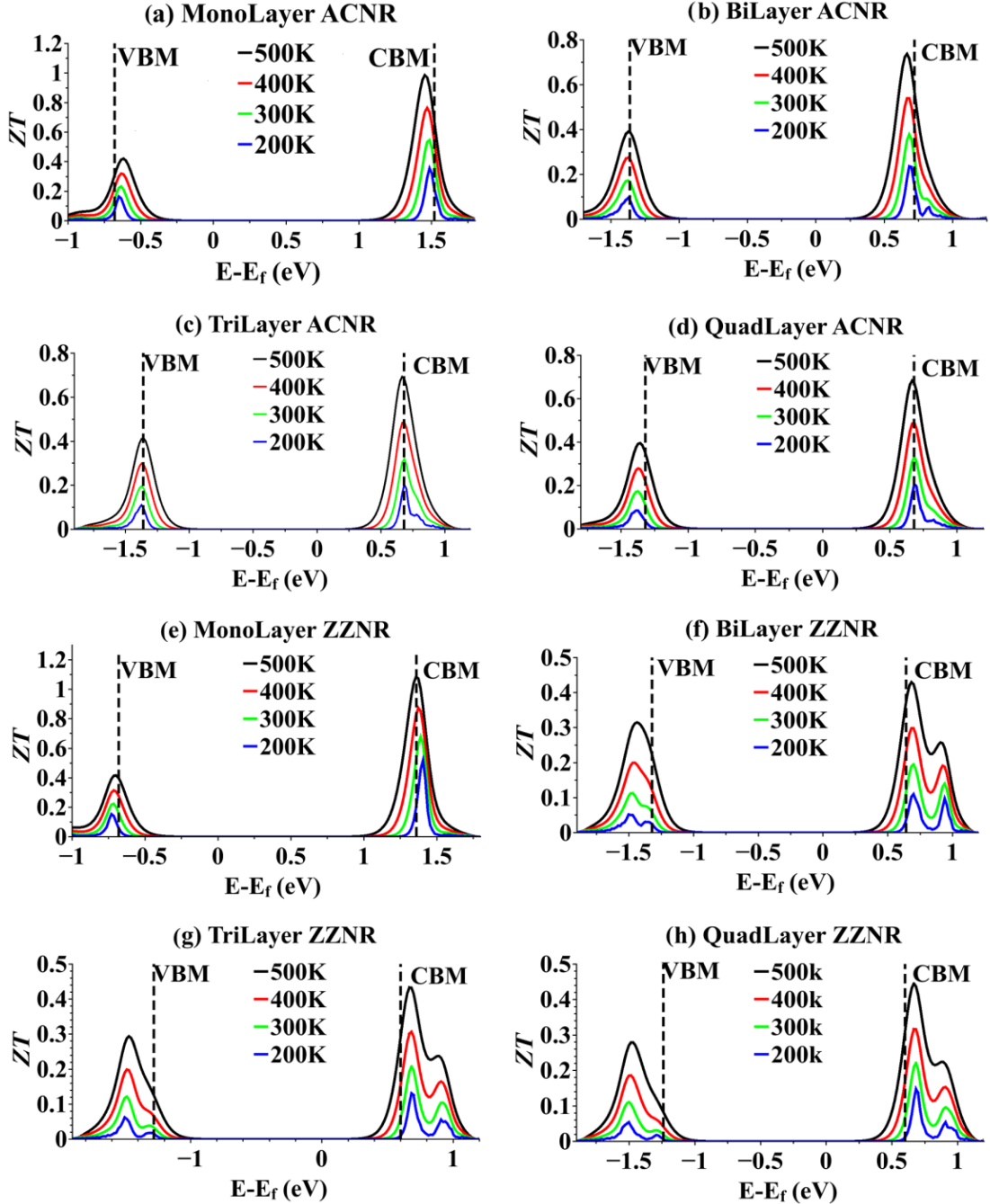

Fig. 6. Thermoelectric figure of merit *ZT* for monolayer and few-layer $MoS_2$ ACNR and ZZNR vs. Fermi level position for four temperatures. Conduction band minimum (CBM) and valence band maximum (VBM) are shown by vertical dashed lines in each plot.

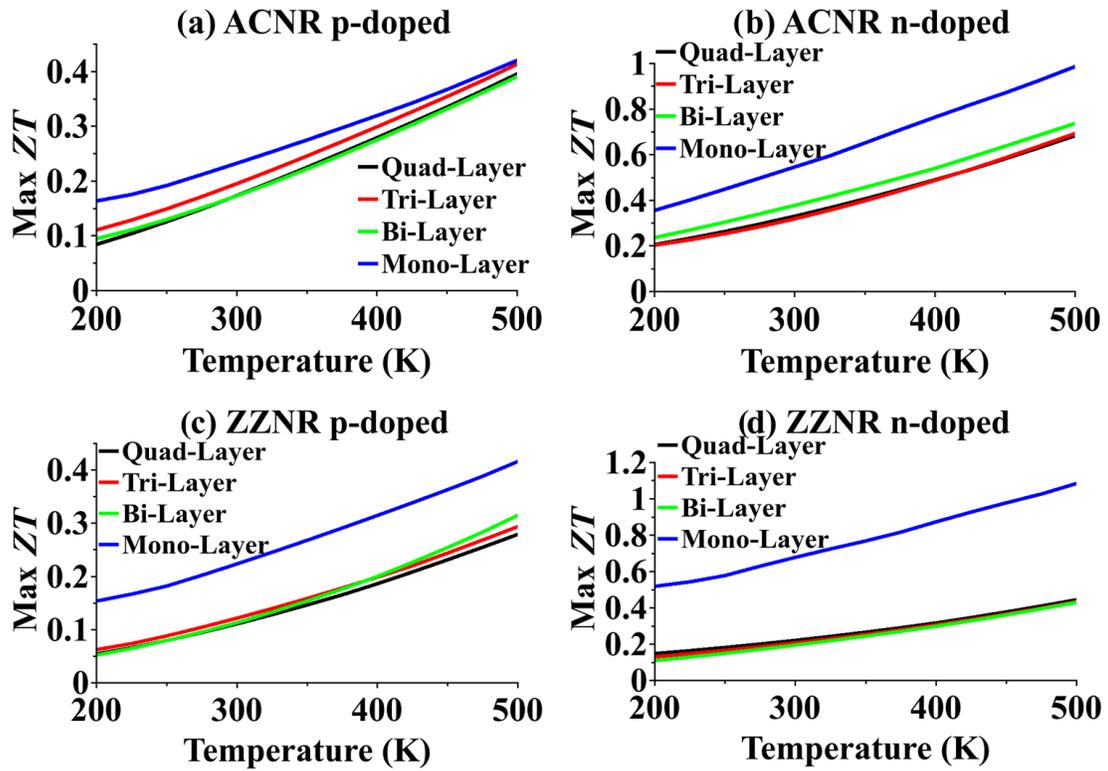

Fig. 7. Max *ZT* for p-doped and n-doped monolayer and few-layer ACNR and ZZNR vs. temperature.

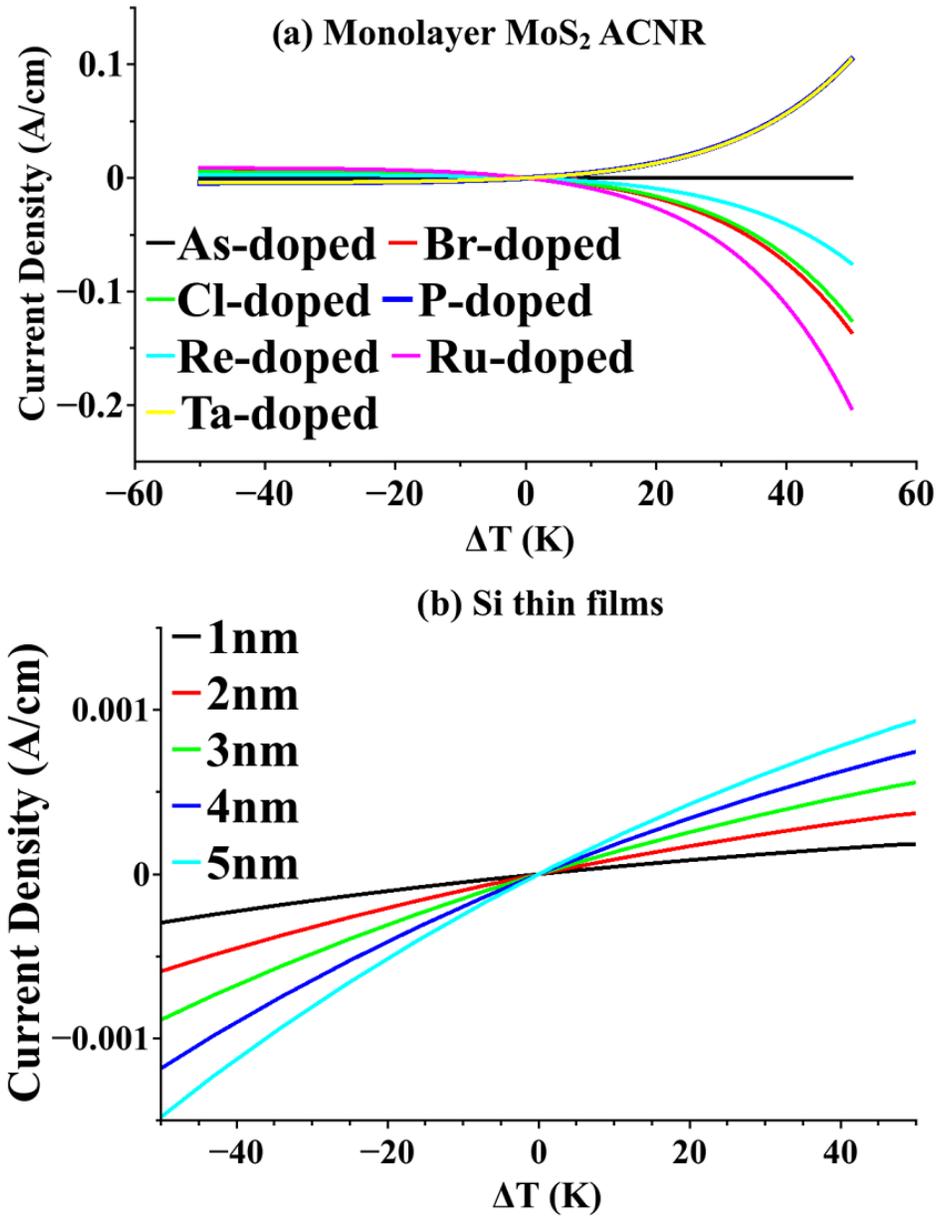

Fig. 8. (a) Thermoelectric current of monolayer $MoS_2$ ACNR substitutionally doped with various dopants vs. temperature gradient across nanoribbons. Transition metal dopants replace Molybdenum and non-metal dopants replace Sulfur. (b) Thermoelectric current of Si thin films doped p-type for different film thicknesses vs. temperature gradient across films.

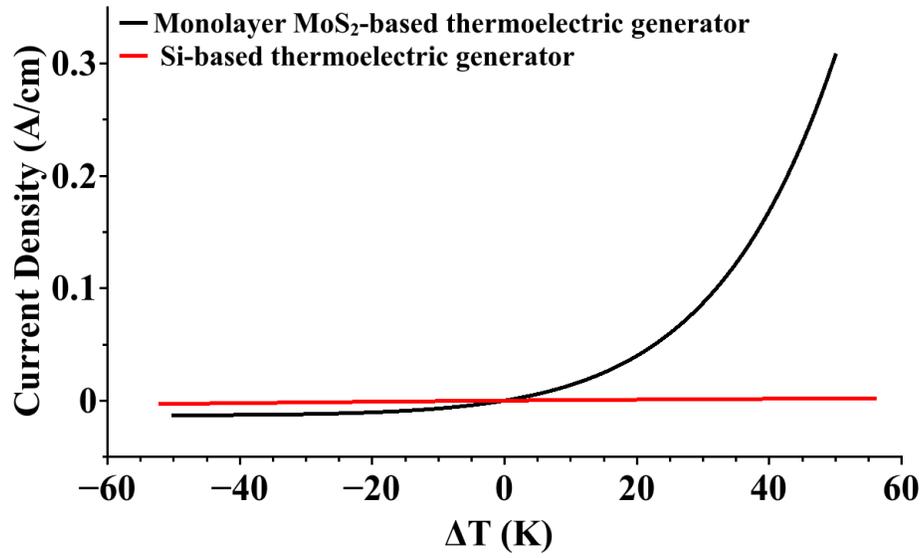

Fig. 9. Thermoelectric current of thermoelectric generators based on Si thin film in comparison with that of based on monolayer $MoS_2$.